\documentclass[
prx,
superscriptaddress,
twocolumn,
longbibliography
]{revtex4-1}

\usepackage{float}
\usepackage{bbm}
\usepackage{epsfig}
\usepackage{epstopdf}
\usepackage{graphicx, subfigure}
\usepackage{pgfplots}
\usetikzlibrary{patterns}
\usepgfplotslibrary{groupplots}
\usepackage{amssymb}
\usepackage{amsmath}
\usepackage{scalefnt}
\usepackage{color}
\usepackage[title]{appendix}
\usepackage{hyperref}
\usepackage{qcircuit}
\usepackage[capitalize]{cleveref}
\usepackage{braket,mleftright}
\usepackage{physics}
\usepackage{dcolumn}
\usepackage{csquotes}
\usepackage[T1]{fontenc}

\begin{document}
\title{A Classically Efficient Quantum Scalable Fermi-Hubbard Benchmark}

\author{Bryan T. Gard}
\email{Bryan.Gard@gtri.gatech.edu}
\affiliation{Georgia Tech Research Institute, Atlanta, GA 30332, USA}
\author{Adam M. Meier}
\affiliation{Georgia Tech Research Institute, Atlanta, GA 30332, USA}
\date{\today}

\begin{abstract}
In order to quantify the relative performance of different testbed quantum computing devices, it is useful to benchmark them using a common protocol. While some benchmarks rely on the performance of random circuits and are generic in nature, here we instead propose and implement a practical, application-based benchmark. In particular, our protocol calculates the energy of the ground state in the single particle subspace of a 1-D Fermi Hubbard model, a problem which is efficient to solve classically. We provide a quantum ansatz for the problem that is provably able to probe the full single particle subspace for a general length 1-D chain and scales efficiently in number of gates and measurements. Finally, we demonstrate and analyze the benchmark performance on superconducting and ion trap testbed hardware from three hardware vendors and with up to 24 qubits.
\end{abstract}

\maketitle

\section{Introduction}
The performance of Noisy Intermediate Scale Quantum (NISQ) devices is impacted by a variety of noise sources \cite{Preskill2018quantumcomputingin}, which limit the sizes of quantum circuits that can be implemented effectively. In order to characterize the generic performance of NISQ devices, benchmarks based on random circuits are widely used \cite{Emerson_2005,PhysRevA.80.012304,PhysRevA.77.012307,PhysRevLett.109.240504,PhysRevA.100.032328,Moll_2018,BlumeKohout2020volumetricframework}. However the driving interest in NISQ devices is their promise for solving practical problems including quantum chemistry \cite{doi:10.1021/acs.chemrev.8b00803,Google_chemsitry}, combinatorial optimization \cite{farhi2014quantum,farhi2019quantum,PhysRevE.99.013304,rajakumar2020generating,martiel2021benchmarking}, quantum simulation \cite{Buluta108,Britton2012}, and machine learning \cite{Biamonte2017,doi:10.1080/00107514.2014.964942}. In order to probe these applications more directly, benchmarks directly mimicking the application have recently arisen~\cite{dallairedemers2020application,karamlou2021analyzing,martiel2021benchmarking}. We highlight one hybrid quantum algorithm that has received significant attention, the Variational Quantum Eigensolver (VQE) \cite{Peruzzo2014,kandala2017hardware,PhysRevLett.122.140504,PhysRevLett.122.230401,gard2019efficient,McCaskey2019qchem,Cade_2020}, which has the potential to solve practical near-term problems that scale inefficiently on classical machines. 

A NISQ application-specific benchmark benefits from several key properties. It should be scalable in the sense that it can be applied to arbitrarily sized quantum systems. For near term implementations, it is also desirable for the benchmark to have low circuit depth.  Similarly, it should require few unique circuits and measurement settings, thus lessening the required run time and cost on commercial testbed devices. The benchmark should be straightforward to implement efficiently on a broad class of hardwares and architectures, which generally favors constructions based on standard one and two-qubit gates with nearest-neighbor connectivity. Finally, the benchmark should discriminate performance between currently available devices. 

In this work, we introduce and demonstrate an efficient application-based benchmark focused on finding low-energy states of the Fermi-Hubbard model, in a similar spirit to work from Dallaire-Demers {\it{et. al.}}~\cite{dallairedemers2020application}. We construct a scalable quantum ansatz for the problem which uses relevant symmetries in the problem to reduce the required quantum resources~\cite{gard2019efficient,barron2020vqe,doi.org/10.1002/qute.202100012,Barkoutsos2018}.

We show that the construction of the quantum circuit for use in the benchmark has a gate count and optimization parameter count which scales linearly in the number of qubits. Additionally, the benchmark also requires only a constant number of non-commuting sets of measurements (independent of problem size). Additionally, our ansatz only assumes nearest-neighbor connectivity and uses a common gate set consisting of CNOT entangling gates and parametrized single-qubit rotations. We also implement the benchmark on nine different testbed devices from three different providers to demonstrate how broadly approachable it is for current hardware.

Any quantum computer benchmark reports the performance of a composite of the quantum hardware and the software required to program the hardware. For each device we benchmark, Qiskit or Amazon Braket software submits quantum circuit jobs to the respective quantum hardware. This software also maps virtual qubits and gates in the quantum circuits to specific physical qubits and gates on the device; this mapping can be optimized by the software to highlight the best performing qubits and interactions. Therefore the resulting performance is a consequence of both the hardware performance and the automated selection done by the software.

The primary result of our benchmark is a single number representing the largest size (twice the number of fermionic sites) of the Fermi-Hubbard calculation implemented in the protocol that returns a result below an error threshold. This output is similar in kind to the ``algorithmic qubits'' benchmark~\cite{ionq_alg} in that it represents the maximum number of qubits in the device that can be effectively utilized to solve a problem. 

We begin in Section~\ref{sec:Benchmark} with a description of our benchmark and its required components. In Section~\ref{sec:Results} we show the results of our benchmark run on IBM, IonQ and Rigetti hardware. We discuss our results and conclude in Section~\ref{sec:Discussion}.

\section{Benchmark}
\label{sec:Benchmark}
While benchmarks like Randomized Benchmarking (RB) and Quantum Volume (QV) attempt to capture generic performance based on random circuits constructed within a rubric, it is also interesting to consider the performance of quantum devices based on their ability to solve practical and specific problems. We focus on the problem of finding the ground state of a physically relevant Hamiltonian, which is a common underlying task in NISQ applications. In particular, we choose the Hamiltonian for the 1-D Fermi-Hubbard model because of its simplicity and familiarity, as well as its symmetry properties which lead to simpler circuits. The general 1-D Fermi-Hubbard model is given by the equation,
\begin{equation}
    \hat{H}=-t\sum_{\langle i,j\rangle, \sigma} (\hat{a}_{i,\sigma}^\dagger\hat{a}_{j,\sigma} +\hat{a}_{j,\sigma}^\dagger\hat{a}_{i,\sigma} ) + U \sum_i \hat{a}_{i,\uparrow}^\dagger\hat{a}_{i,\uparrow}\hat{a}_{i,\downarrow}^\dagger\hat{a}_{i,\downarrow},
\label{eq:FH}
\end{equation}
where $\hat{a}_{i,\sigma}^\dagger$ ($\hat{a}_{i,\sigma}$) is the creation (annihilation) operator associated with site $i$ and spin $\sigma$. We map this fermionic problem to a qubit problem by way of the Jordan-Wigner mapping, which directly maps spin orbitals to qubits while maintaining the anti-commutation rules of the original Hamiltonian~\cite{PhysRevLett.63.322}. This maps a problem defined for a $L$-site chain to $N=2L$ qubits (since each site has two spins). 

Solving for the ground state of the generic Fermi-Hubbard Hamiltonian is computationally hard. A brute force classical solution for a 2-D grid based Fermi-Hubbard model requires diagonalization of a size $2^{2n_xn_y} \times 2^{2n_xn_y}$ matrix, for $n_x,n_y$ number of sites in the horizontal and vertical direction~\cite{PhysRevB.102.235122}. The model is analytically solvable in the limiting cases of $\frac{U}{t}\rightarrow0$ and $\frac{U}{t}\rightarrow\infty$ and in the 1-D case using the Bethe ansatz, but general exact solutions are not known~\cite{LIEB20031,doi:10.1142/2148,vandyke2021preparing}. Here, to further simplify the problem for current NISQ hardware and to make a more intuitive benchmark, we restrict our focus to the single particle ground state. In this regime, we can take $U=0$ without loss of generality because, for the single particle ground state, there are no possible interactions, and therefore Eq.~\ref{eq:FH} is invariant to the choice of $U$. This choice leads to a simpler form for the Hamiltonian that can be solved exactly \cite{OSIPOV2017173}. We further fix $t=1$ and choose open (non-periodic) boundary conditions. 

Under these conditions, the analytic, single-particle ground state energy of the 1-D Fermi Hubbard model is,
\begin{equation}
    E_{gs}=2\cos[L\pi/(L+1)],
\label{eq:gs_energy}
\end{equation}
where $L$ is the number of sites in the chain. This simple solution for the restricted problem gives an easy target for the benchmark at any length $L$. Note that considering $U=0$ was only a tool to clarify the ideal exact solution; for measurement of energies on hardware, states which are not only single particles are present due to noise. In order to capture these errors in the benchmark results and tie the benchmark implementation to the more general problem, we use the full hamiltonian with $U=2$. 

To test the ability of NISQ devices to find this target ground state, we employ the VQE algorithm. This algorithm relies on the fact that, for any parameterized quantum state $\psi(\theta_i)$, minimizing $\langle \psi(\theta_i)|H|\psi(\theta_i) \rangle$ over $\theta_i$, bounds the ground state energy.  If the parameterization allows for the creation of exactly the space of states allowed by the problem constraints, e.g., fixed particle number, then optimizing the energy in this way solves for a ground state subject to those constraints (there may be many due to degeneracy). 

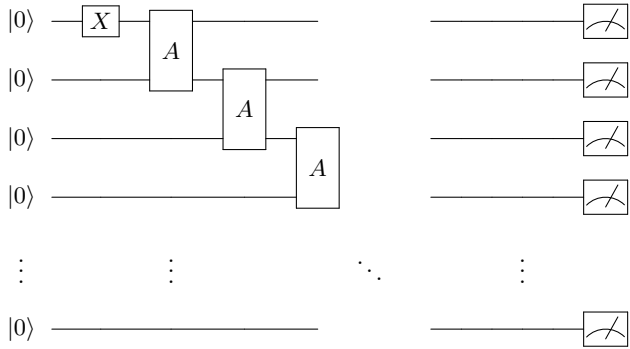
\begin{figure}[!tb]
            \[ \Qcircuit @C=1.25em @R=1em {
            \ket{0} &	&	\gate{X}	&	\multigate{1}{A}	&	\qw	&	\qw	&		&		&		&	\qw	&	\qw	&	\qw	&	\qw & \meter	\\
            \ket{0} &	&	\qw        	&	\ghost{A}	&	\multigate{1}{A}	&	\qw	&		&		&		&	\qw	&	\qw	&	\qw	&	\qw & \meter	\\
            \ket{0} &	&	\qw	&	\qw	&	\ghost{A}	&	\multigate{1}{A}	&		&		&		&	\qw	&	\qw	&	\qw	&	\qw	 & \meter\\
            \ket{0} &	&	\qw	&	\qw	&	\qw	&	\ghost{A}	&		&		&		&	\qw	&	\qw	&	\qw	&	\qw	 & \meter\\
	        &		&		&		&		&		&		&		&		&		&		&		 \\
            \vdots	&		&		&	\vdots	&		&		&	\ddots	&		&		&		&		&	\vdots	 \\
	        &		&		&		&		&		&		&		&		&		&		&		\\
            \ket{0} &	&	\qw	&	\qw	&	\qw	&	\qw	&		&		&		&	\qw	&	\qw	&	\qw	&	\qw	 & \meter
            } \]
            \caption{Circuit used to produce the single particle ansatz. Each $A$ gate preserves particle number and spin projection. $A$ gates are present on the top half of qubits and measurements are taken on all qubits.
            \label{fig:general_ansatz}}
\end{figure}

To implement the VQE algorithm we also need to specify an ansatz (sometimes also called a variational quantum circuit). For resource efficiency, we choose an ansatz which is built to enforce the symmetries in the Hamiltonian. Ans\"{a}tze of this kind have been used in previous works related to symmetry preserving circuits for chemical ground states~\cite{PhysRevA.98.022322,gard2019efficient,barron2020vqe,PhysRevB.102.075104,vandyke2021preparing}. Our ansatz, depicted in Fig.~\ref{fig:general_ansatz}, is built from a single primitive gate, $A$, with a single parameter, $\theta$, where $A$ has the following computational-basis matrix representation:
 \begin{eqnarray}
                A(\theta)=\begin{pmatrix}
                1 & 0 & 0 & 0\\
                0 & \sin \theta  & \cos \theta  & 0\\
                0 & \cos \theta & -\sin \theta & 0\\
                0 & 0 & 0 &1
            \end{pmatrix}.\label{eq:Agate}
 \end{eqnarray}
            
Constructing an ansatz from this gate has several beneficial features. First, an ansatz built from parameterized SWAP-type gates naturally conserves particle number. In addition, when Eq.~\ref{eq:FH} is mapped to qubit operators using a spin-block based Jordan-Wigner mapping, our ansatz also preserves spin projection. The ansatz also maps real states to real states, and we note that all ground states of our chosen problem are real valued. Since these symmetries are encoded into the ansatz, it fundamentally requires fewer quantum resources than a more general ansatz, e.g. Ref.~\cite{kandala2017hardware,rattew2020domainagnostic}. Since the problem of interest is a single particle ground state, we start with a single excited qubit and construct a simple ladder-like circuit for an arbitrary number of sites, as shown in Fig.~\ref{fig:general_ansatz}.
Intuitively, the proposed circuit can swap the initial single particle excitation into any qubit, controlled by the parameters $\theta_i$ for qubit $i$. 

\begin{figure}[!tb]
           \[ \Qcircuit @C=1em @R=0.75em {
&	\multigate{2}{A(\theta)}	&	\qw	&		&		&		&	\targ	&	\qw	&	\ctrl{+2}	&	\qw	&	\targ	&	\qw	\\
&		&		&		&	=	&		&		&		&		&		&		&		\\
&	\ghost{A(\theta)}	&	\qw	&		&		&		&	\ctrl{-2}	&	\gate{R_y(\theta)^\dagger}	&	\targ	&	\gate{R_y(\theta)}	&	\ctrl{-2}	&	\qw
} \]
\caption{Decomposition of the $A$ gate in terms of elementary single and two-qubit gates. $R_{y}(\phi )=\exp (-i\phi \sigma _{y}/2)$.}
\label{fig:Adecomp}
\end{figure}
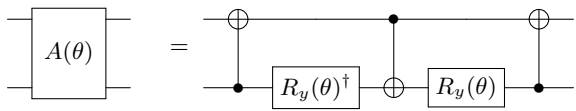

In the standard gate set for quantum computing and for a generic input state, each $A$ gate can be decomposed into 3 CNOT and 2 single-qubit gates, as shown in Fig.~\ref{fig:Adecomp}.
However, since we always begin the benchmark circuit in a fixed initial state, we can make small simplifications to the general circuit, as shown in Figs.~\ref{fig:Asimp1} and \ref{fig:Asimp2} in Appendix~\ref{app:Asimp}.

Based on these simplifications, the ansatz requires $2L-3$ CNOT gates and $L-1$ parameters for $L\geq 2$, the number of sites in the Fermi-Hubbard chain. Since the ansatz only requires 1-D nearest-neighbor connectivity, this gate count can be achieved by any device with nearest neighbor CNOT gates, avoiding the overhead frequently required to map arbitrarily connected circuits to a specific hardware connectivity. However, the benchmark also does not take advantage of arbitrary connectivity present in some quantum systems such as ion traps. Finally, we prove in Appendix~\ref{app:exact} that the ansatz can identify the single particle ground state for any $L$.

\begin{figure}
\includegraphics[scale=0.4]{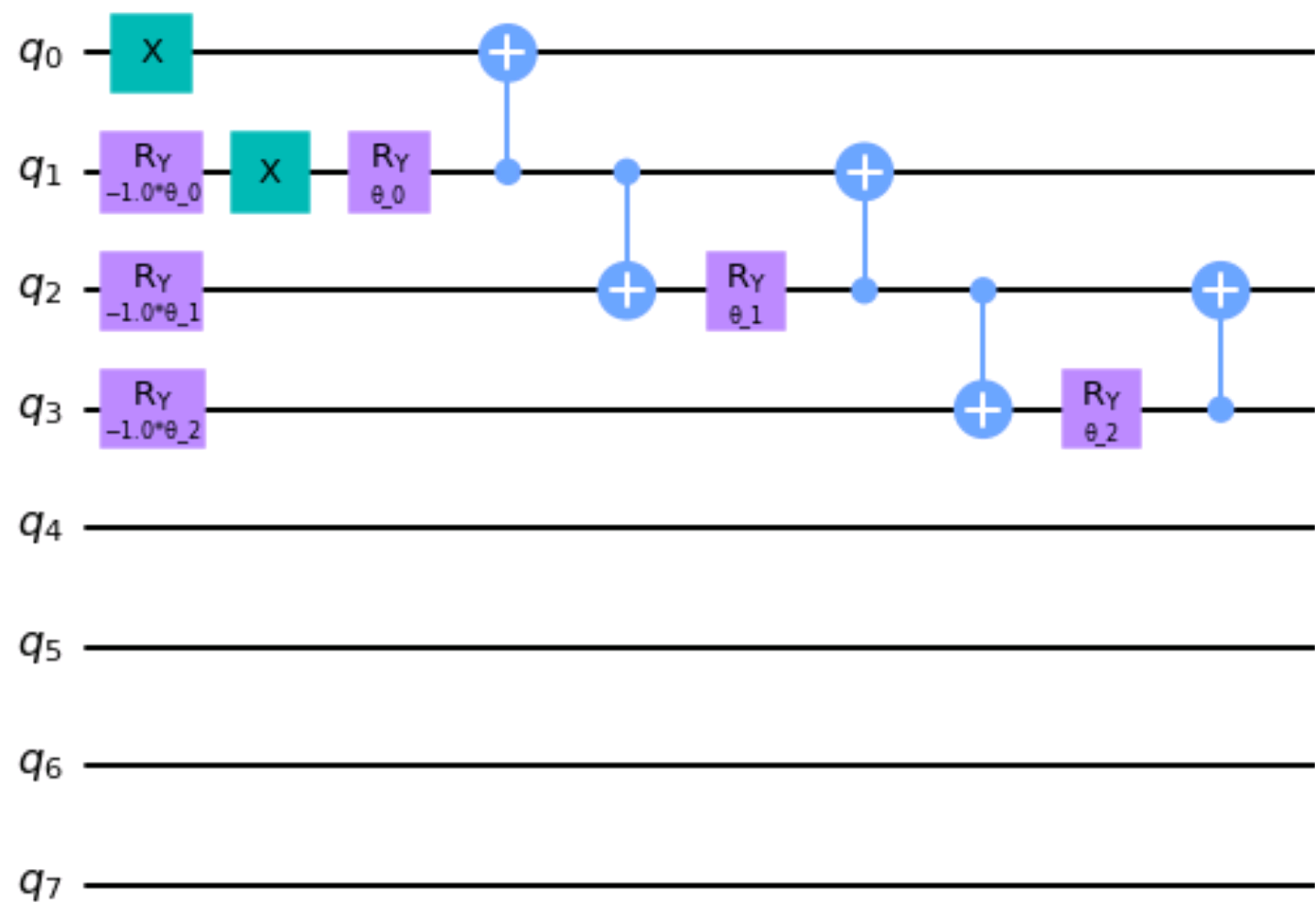}
\caption{Example symmetry based ansatz for $L=4$, composed of three simplified $A$ gates and three parameters. While gates are only required on the first $L$ qubits, measurement of all $N=2L$ qubits is performed to calculate the energy of the Hamiltonian defined on $N$ qubits.}
\label{fig:L4example}
\end{figure}

Fig.~\ref{fig:L4example} presents an example of the ansatze for $L=4$. Notably, the circuit only fills the top half of the quantum register. This is a direct consequence of the single particle ground state necessarily having a spin projection value of $s_z=\pm0.5$. Because we have mapped the original fermionic problem to qubit operators using the Jordan-Wigner mapping, and we have used a block-based spin encoding, the top half of qubits encodes spin up particles, while the remaining qubits encode spin down. The two choices of $s_z$ are degenerate with one another and are related by a simple bit flip operation on all qubits, so we need only generate one of these choices, and we choose the $s_z=0.5$ state. Note that the general Hamiltonian itself requires $2L$ qubits to express, even though gates are only present on the first $L$ qubits in our restricted problem. In our benchmark, we still choose to measure all $2L$ qubits to calculate the energy of the single particle state, and this choice means that measurement errors on the unused qubits still impact the result. In principle we could allow devices which employ mid-circuit reset to reset the empty qubits prior to measurement, but we do not use this in our benchmark. Writing the protocol in this way preserves our ability to extend the Hamiltonian to more than one particle without changing the construction. We further discuss extension of the benchmark to more particles in Appendix~\ref{app:extend}.

The problem Hamiltonian, once mapped to qubit operators, can be written as a sum of Pauli strings. For instance, the Hamiltonian of the simplest case at $L=2$ is
\begin{eqnarray}
\hat{H}=&IIII+\frac{1}{2}(IZIZ+ZIZI-IZYY-IZII-ZIII \nonumber \\
&-IZXX-YYZI-IIIZ-IIZI-XXZI)\nonumber.
\end{eqnarray}

For most devices, measuring any operators that contain $X$ or $Y$ requires that we rotate the qubits prior to a $Z$ basis measurement. We can also group these strings into a small number of sets where all strings within each set commute with one another and therefore can be measured simultaneously. In this example case, we can group the Hamiltonian into five groups of terms given by,
\begin{eqnarray}
\hat{H}=&H_0+H_1+H_2+H_3+H_4\nonumber \\
H_0=& IIII+\frac{1}{2}(IZIZ+ZIZI-ZIII\nonumber \\
&-IZII-IIZI-IIIZ)\nonumber \\
H_1=&-\frac{1}{2}IZXX\nonumber\\
H_2=&-\frac{1}{2}IZYY\nonumber\\
H_3=&-\frac{1}{2}YYZI\nonumber\\
H_4=&-\frac{1}{2}XXZI\nonumber
\end{eqnarray}
Note that this grouping is not unique and many other choices exist but do not change the resulting measurement of $\langle \hat{H} \rangle$. In the general case, the Fermi-Hubbard Hamiltonian can \textit{always} be measured with at most five commuting sets, independent of the problem size~\cite{PhysRevB.102.235122}. This, coupled with the linear scaling of qubits and gates for the ansatz, makes the benchmark both simple to implement at small sizes and simple to scale to larger sizes.

The motivating application for our benchmark is the use of the VQE hybrid algorithm in order to find minimum energy solutions on quantum hardware. To find the minimum energies, the algorithm optimizes a parameterized circuit (creating a parameterized state). While the VQE algorithm can in principle start with any seed parameters, we can assist it by pre-optimizing the parameters classically.  In the case of the hardware devices tested, we do not actually perform the hybrid algorithm at all. Instead we only evaluate the energy of the pre-optimized parameters. This choice reduces the computational (and monetary) cost of the benchmark, which is only formally dependent on the accuracy of the energy of the optimized parameters. Through comparison of the energy associated with the pre-optimized parameters with the ideal result, we can show that the pre-optimization does not appreciably impact the benchmark results: the maximum energy difference we find in this comparison is $2.946 \times 10^{-10}$ at ($L=12$). This demonstrates that the VQE algorithm itself along with classical optimization is able to very accurately find the desired minimal energy states and does not limit the benchmark results we report.

\begin{figure}
    \centering
    \includegraphics[scale=0.2]{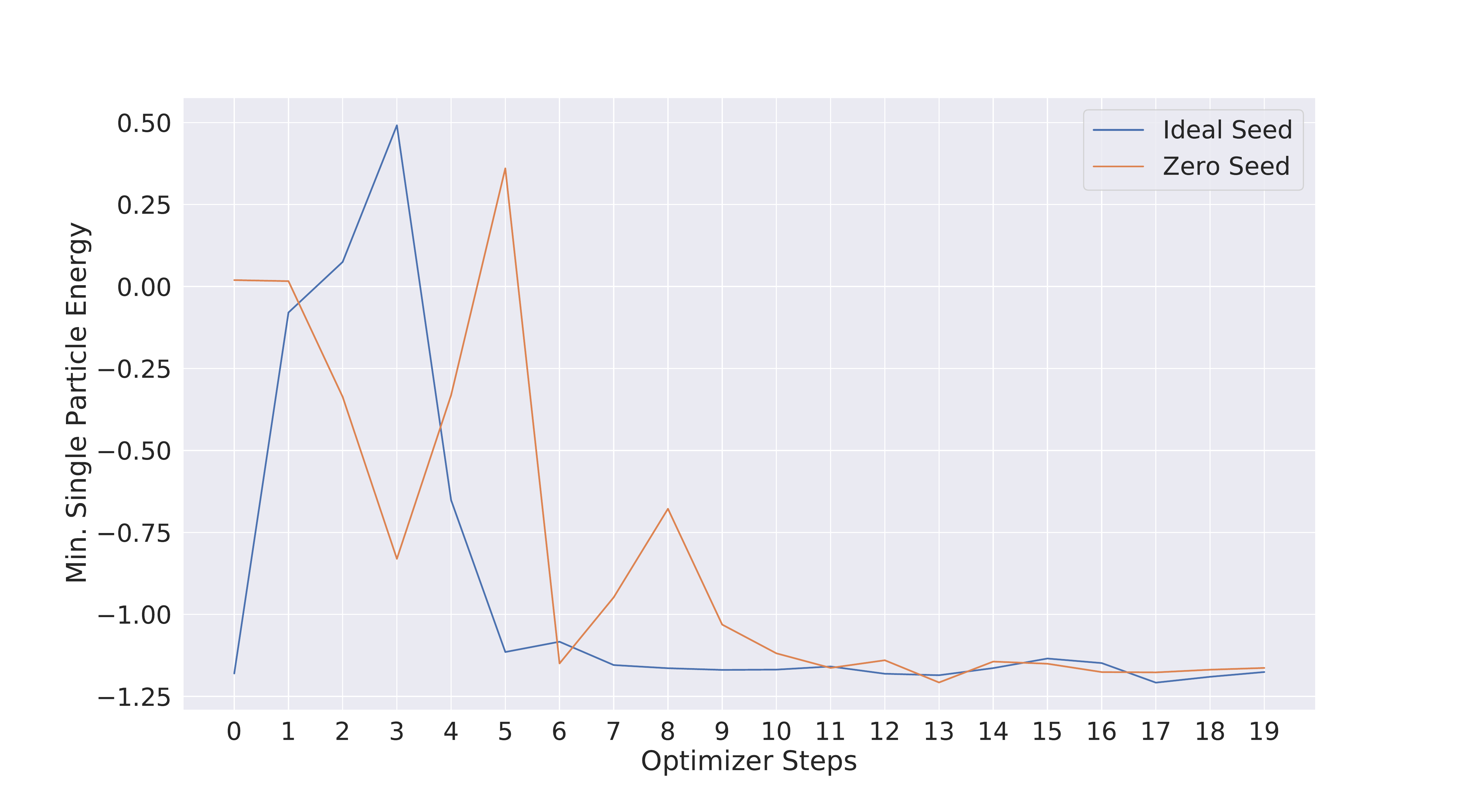}
    \caption{VQE convergence using the COBYLA optimizer over 20 steps on IBM Guadalupe hardware, using an initial point seeded from an ideal simulation and using a fixed initial point of [0,0].}
    \label{fig:opt_steps}
\end{figure}

In Fig.~\ref{fig:opt_steps} we show an example of the convergence behavior of the full VQE algorithm over 20 steps using IBM Guadalupe for the 3-site Fermi Hubbard benchmark. We plot the convergence for two different sets of initial parameters: one with initial parameters arbitrarily fixed to zero and the second with the classically pre-optimized parameters. The use of the ideal parameters assists by improving the minimum number of steps before convergence of the optimizer but is not necessary. This simple case with only two parameters indicates the expected result that the VQE algorithm is able to converge on hardware eventually without the need of any specific initial point derived from simulation. 

The issue of how to optimize hybrid performance with respect to splitting resources between VQE evaluations and classical pre-optimization for large application instances is an interesting and important problem. Performing poorly at this optimization could impact very large instances of our Fermi length benchmark and is an example of how the benchmark evaluates both hardware and software aspects of NISQ devices; however, the instances here are too small for the difficulty of classical optimization to have an impact.

\section{Results}
\label{sec:Results}
We run the benchmark described in the previous section on nine devices provided by IBM, Rigetti Computing, and IonQ. We accessed private IBM devices (7 total) through IBM Qiskit software \cite{Qiskit}. We accessed IonQ and Rigetti devoces through Amazon Web Services' BraKet software. 

For each choice of chain length, we seed the hardware evaluations by first running a simulator of the VQE algorithm using the high-performance Qulacs software package~\cite{qulacs}. These simulations provide optimal parameters $\theta_i$ for each choice of $L$ that we use to initialize the same problem on the quantum devices. For all hardware, we  evaluate the energy at the fixed seed parameter values in order to reduce cost. The benchmark evaluation relies only on the energy evaluation of this single point for each $L$. The Qulacs classical simulation is not a required step for the actual VQE algorithm; in principle, the same results can be obtained by implementing more steps of the hybrid algorithm on the quantum hardware at a greater cost to the quantum devices.

\begin{figure*}[htb!]
\includegraphics[trim=20 20 20 20, clip,scale =0.35]{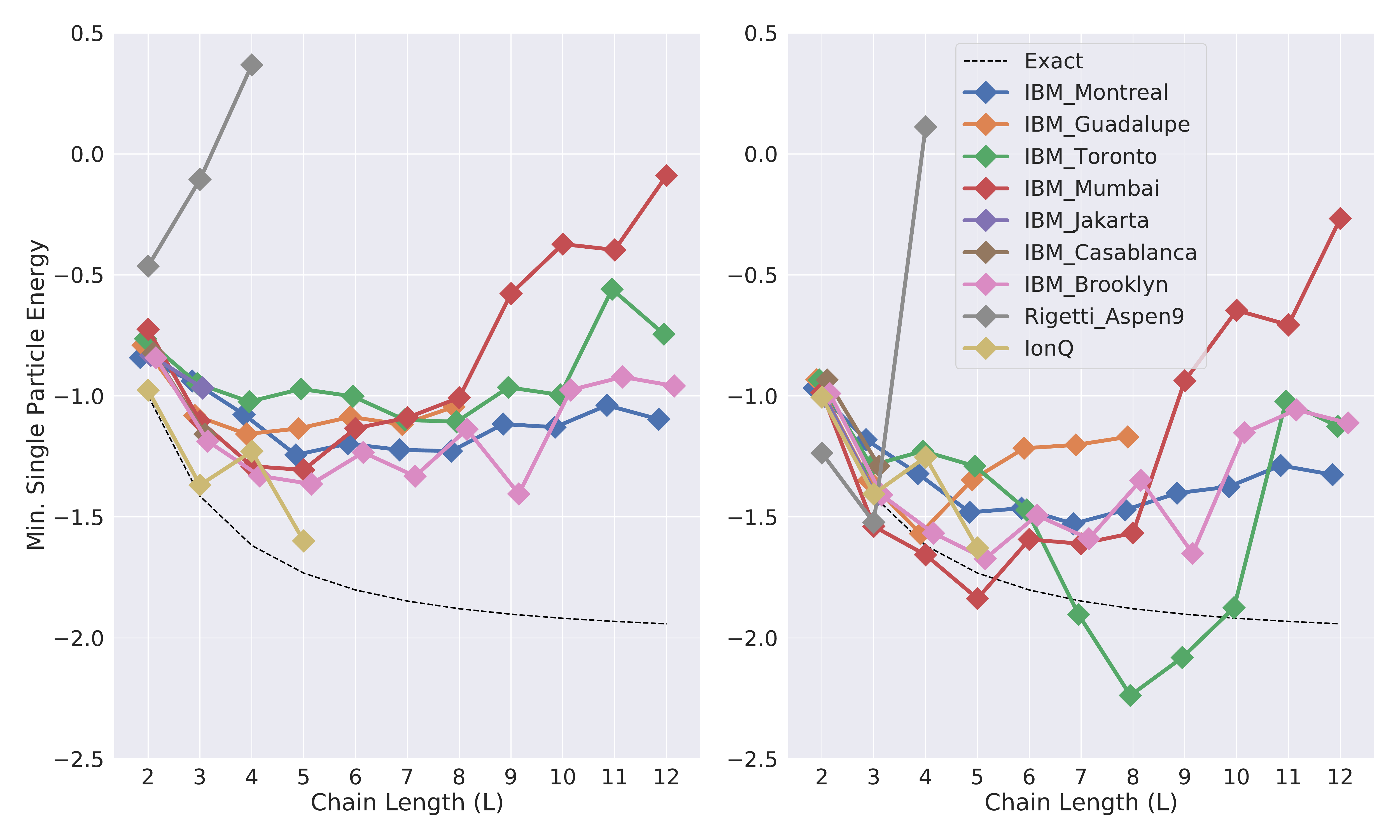}
\caption{(Left) Raw data of the single particle energy as a function of Fermi-Hubbard chain length ($N=2L$). In all cases, error bars are smaller than marker size. (Right) Mitigated data of the single particle energy after SPAM error mitigation. Data points are horizontally offset slightly from their integer values to reduce clutter.}
\label{fig:FH_raw}
\end{figure*}

For each device, we run the benchmark for increasing chain lengths starting with $L=2$ and terminating when the benchmark performance is severely degraded or when the maximum system size is reached. For all hardware devices, we fixed the number of repetitions (shots) to 8192. In Table~\ref{tbl:devices} we capture the qubit number, the quantum volume reported elsewhere in literature, and our benchmark result for each device. Of the nine devices we benchmarked, five were able to run to their full physical limit, while the remaining four were terminated early (note that the protocol requires an even number of qubits).

We show the raw energies computed using each device in Fig.~\ref{fig:FH_raw}. In the right panel, we show the raw results and we observe that the superconducting devices typically have larger errors than the IonQ device, which performs well through $L=5$ (with the exception of $L=4$). Most devices visually follow the trend of the ideal energy at short chain lengths before crossing a threshold and drifting further away from the exact energy. This makes the benchmark easy to evaluate qualitatively. However, either due to noise in hardware or differences in software optimization, the threshold can be difficult to identify exactly, e.g., for the IBM Brooklyn device which performs better at length 9 than at length 8. For this reason we establish in the following paragraphs a criterion that condenses the benchmark results to a single number for report.

\begin{table}[h!]
    \centering
    \begin{tabular}{|c|c|c|c|c|}
       \hline Device & Max. Qubits & Log(QV) & $L^*$ (R) &  $L^*$ (M)  \\
        \hline Jakarta & 7 & 4 & 4* & 6\\
        \hline Casablanca & 7 & 5 & 6 & 6\\
        \hline Guadalupe & 16 & 5 & 0* & 10*\\
        \hline Toronto & 27 & 5 & 0* & 6*\\
        \hline Mumbai & 27 & 7 & 0* & 16*\\
        \hline Montreal & 27 & 7  & 4* & 24\\
        \hline Brooklyn & 65 & 5  & 10* & 24\\
        \hline IonQ & 11 & -  & 6* & 10\\
        \hline Rigetti Aspen9 & 32 & -  & 0* & 0*\\
        \hline
    \end{tabular}
    \caption{Table of devices used in our benchmark along with their maximum usable qubits, quantum volume (where available) and the result of our benchmark. $L^*$ lists the maximum number of qubits (twice the chain length) used which pass the benchmark for both the raw (R.) and mitigated (M.) results. A symbol of (*) denotes benchmarks which were terminated because they exceeded the threshold error score, otherwise the benchmark is run up to the maximum size allowed by each device or post processing. We used $10^{-3}$ as our error score threshold for passing the benchmark as described in the body text.}
    \label{tbl:devices}
\end{table}


We find that the explanatory power of the benchmark results benefits from the option of including standardized State Preparation And Measurement (SPAM) error mitigation in the protocol. We describe our approach to SPAM mitigation in Appendix~\ref{app:symmerr}. SPAM mitigation leads to significantly better energy estimation across most devices tested. The results of applying the SPAM mitigation are shown in right panel of Fig.~\ref{fig:FH_raw}, and here the hardware results track the ideal results more closely than in the left panel, though significant errors are still present. When characterizing the performance of a device with our benchmark, we consider the performance obtained after SPAM mitigation to be the primary result. However, we note that this mitigation technique can require significant classical computing resources for large problem sizes. For this reason, we only include benchmark results up to 24 qubits. We also present the raw (unmitigated) results in both the figure and the table of results for comparison.

\begin{figure*}[!htb]
\includegraphics[scale=0.3]{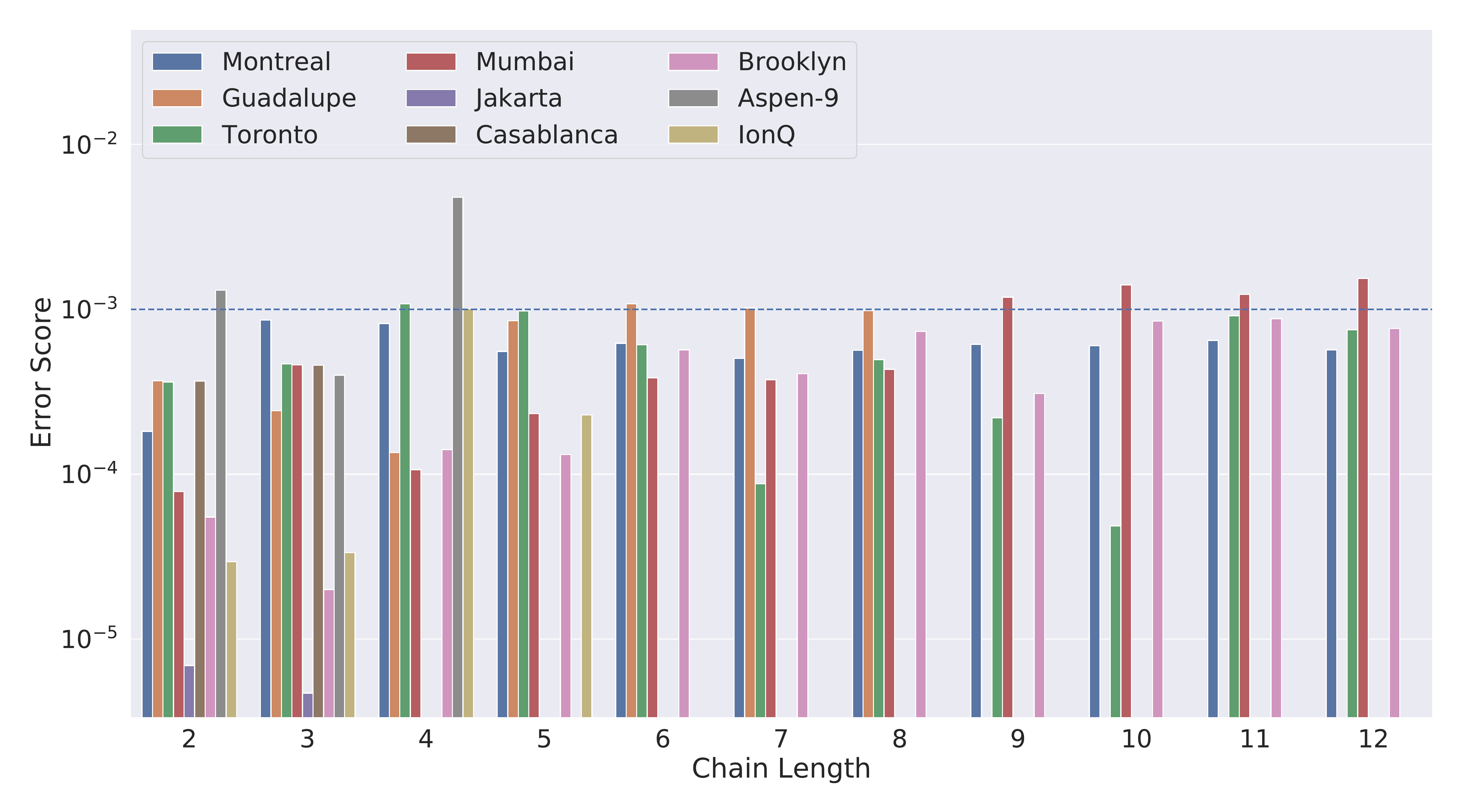}
\caption{Error score of each device, grouped by chain length (lower error score indicates a better result). An error threshold (dashed) is shown at a value of $10^{-3}$. For each device, when the first violation of this threshold is found, the device is said to have failed the benchmark.}
\label{fig:FH_diff}
\end{figure*}

In order to distill the benchmark to a single number, we choose a benchmark score inspired by the accuracy requirement in the LINPACK benchmark~\cite{linpack}. We define our error score by:
\begin{equation}
    \mathrm{Error}~\mathrm{ Score} = \frac{1}{\sqrt{M}}\times|\langle E\rangle -E_{gs}|/L ,
\end{equation}
where $\langle E \rangle$ is the measured energy from the quantum device, after SPAM mitigation, $E_{gs}$ is the known classical result, $L$ is the chain length, and $M=8192$ is the number of shots in each measurement. As we run our benchmark for increasing $L$,  we note the first length occurrence $L^*+1$ for which each device has an error score greater than $10^{-3}$, and we report that the device passes the benchmark up to a length $L^*$. The threshold value of $10^{-3}$ was chosen arbitrarily. It may be useful to consider other threshold values in the future, however this alteration of the benchmark should be reported clearly for the sake of fair comparison. We capture the error score for each device and chain length in Fig.~\ref{fig:FH_diff}. Here we can clearly determine when each device exceeds our chosen threshold.

According to this metric, IBM's Montreal device, which has one of the largest quantum volume of all devices we tested, also performs very well in our benchmark, based on a combination of large qubit number and low noise. The Montreal device results improved significantly upon applying SPAM mitigation, but others, e.g., IonQ, improve less noticeably. This is expected as the IonQ device typically has significantly smaller SPAM errors than the superconducting devices. See Appendix~\ref{app:symmerr} for further details regarding SPAM mitigation. We also wish to highlight the smaller IBM devices, Jakarta and Casablanca, which perform well within this benchmark up to their maximum allowed size. We suggest that the lacking performance of the Rigetti Aspen-9 device is due to errors which are not corrected through simple SPAM mitigation techniques and would require more rigorous error mitigation strategies. 

\section{Conclusion}
\label{sec:Discussion}
We have proposed and demonstrated an application-specific quantum benchmark that is well suited for current and near-term NISQ devices. 
We have shown that our proposed benchmark scales in small increments of the required quantum resources, giving it the ability to discriminate devices finely. The choice to restrict the problem to a single particle subspace of the 1-D Fermi Hubbard model also allows us to exactly solve for the energy at arbitrary $L$, making the benchmark easy to verify classically at any size.

We applied a well-conditioned symmetry based ansatz and SPAM error mitigation strategy in order to improve the results of energy estimation. This ansatz is hardware agnostic and easily defined for arbitrary $L$, and we show analytically that it is always able to find the single particle ground state of the supplied 1-D Fermi Hubbarrd model. The primary result of our chosen benchmark clearly discriminates the performance of available quantum testbeds in a way that tracks closely, but not exactly, with the more abstract ``algorithmic qubits'' metric. Analyses of the impact of SPAM error mitigation and the scaling of the ansatz performance as a function of length provide some clarity about the primary limitations of the devices. The result is a benchmark that is simple to implement and analyze and yet provides good quantitative discrimination between current-generation devices for a problem similar to a commercially-relevant NISQ application.

\section{Acknowledgement}
This research used resources of the Oak Ridge Leadership Computing Facility at the Oak Ridge National Laboratory, which is supported by the Office of Science of the U.S. Department of Energy under Contract No. DE-AC05-00OR22725. We thank John S. Van Dyke and George S. Barron for helpful discussions on the single particle Fermi Hubbard problem.
\bibliography{bib}

\appendix

\section{$A$ Gate Simplification}
\label{app:Asimp}
\begin{figure}[!tb]
           \[ \Qcircuit @C=1em @R=0.75em {
&	\multigate{2}{A(\theta_0)}	&	\qw	&		&		&		&	\qw	&	\qw	&	\qw	&	\qw	&	\targ	&	\qw	\\
&		&		&		&	=	&		&		&		&		&		&		&		\\
&	\ghost{A(\theta_0)}	&	\qw	&		&		&		&	\qw	&	\gate{R_y(\theta_0)^\dagger}	&	\gate{X}	&	\gate{R_y(\theta_0)}	&	\ctrl{-2}	&	\qw
} \]
\caption{Since the initial state is fixed, we can simplify the first $A$ gate (the top most) in the general circuit shown in Fig.~\ref{fig:general_ansatz}. }
\label{fig:Asimp1}
\end{figure}
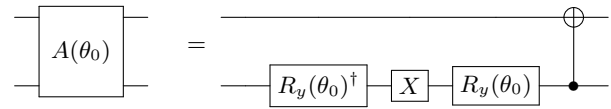

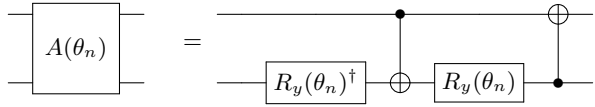
\begin{figure}[!tb]
           \[ \Qcircuit @C=1em @R=0.75em {
&	\multigate{2}{A(\theta_n)}	&	\qw	&		&		&		&	\qw	&	\qw	&	\ctrl{+2}	&	\qw	&	\targ	&	\qw	\\
&		&		&		&	=	&		&		&		&		&		&		&		\\
&	\ghost{A(\theta_n)}	&	\qw	&		&		&		&	\qw	&	\gate{R_y(\theta_n)^\dagger}	&	\targ	&	\gate{R_y(\theta_n)}	&	\ctrl{-2}	&	\qw
} \]
\caption{Simplification of all $A$ ($n>0$) gates in the general circuit except for the first gate, which follows the simplification shown in Fig.~\ref{fig:Asimp1}.}
\label{fig:Asimp2}
\end{figure}

Our efficient ansatz is constructed from a single $X$ gate and many primitive $A$ gates. Due to the construction of the circuit, we can always simplify the general decomposition of the $A$ gate presented in the main text without changing the overall calculation implemented by the circuit. Since the circuit always begins with an $X$ gate on the first qubit, we can simplify the first $A$ gate as shown in Fig.~\ref{fig:Asimp1}. We can simplify all other $A$ gates as shown in Fig.~\ref{fig:Asimp2}. These simplifications are significant improvements for current NISQ devices, which are heavily limited by two-qubit gate error rates.

\section{Exact Ansatz}
\label{app:exact}
In the main text we have used an ansatz that is composed of particle conserving gates that can be decomposed into two or less CNOT gates for the fixed input state we have considered. It is straightforward to show that the ansatz described in the main text is always able to produce any single particle state, and, consequently, that optimization over the parameters of the ansatz can always find the single-particle ground state energy. Consider the first application of an $A$ gate onto the fixed input state defined on $n$ qubits. This first interaction results in the state (up to a global phase),
\[
 A_{0,1}\ket{10\cdots0}=\sin{\theta_0}\ket{10\cdots0}+\cos{\theta_0}\ket{01\cdots0}\equiv \ket{s_0},
\]
where $A_{i,i+1}$ is the gate acting on qubits $i,i+1$ and has corresponding parameter $\theta_i$. The second $A$ gate has a similar action and produces the state
\begin{eqnarray}
 A_{1,2}\ket{s_0}=&\sin{\theta_0}\ket{100\cdots 0}\nonumber\\
 +&\cos{\theta_0}\sin{\theta_1}\ket{010\cdots 0}\nonumber\\
 +&\cos{\theta_0}\cos{\theta_1}\ket{001\cdots 0}\nonumber.
\end{eqnarray}
For the general case of our ansatz on $n$ qubits, we can compactly write the single particle  (in lexicographical order) coefficients as
\begin{eqnarray}
c_0=&\sin{\theta_0}\nonumber\\
c_1=&\cos{\theta_0}\sin{\theta_1}\nonumber\\
c_2=&\cos{\theta_0}\cos{\theta_1}\sin{\theta_2}\nonumber\\
\vdots\nonumber\\
c_{n-1}=&\cos{\theta_0}\cdots\cos{\theta_{n-1}}\sin{\theta_n}\nonumber\\
c_n=&\cos{\theta_0}\cdots\cos{\theta_{n-1}}\cos{\theta_n}\nonumber,
\end{eqnarray}
for $n\geq2$. We can see that repeated applications of the ladder of $A$ gates in the ansatz leads to a state whose coefficients have the form of the coordinates of an $n$-dimensional hypersphere. Since this construction can create any valid hyperspherical coordinates, the ansatz is able to specify any superposition of single particle states with real coefficients.  This set of states is isomorphic to the target (restricted) subspace for our Fermi-Hubbard problem, and so we guarantee that a correct choice of $\theta_i$ is capable of creating the ground state of the subspace. Assuming the optimizer software works in an idealized way, it will be capable of finding this ground state and the corresponding minimum energy.

\section{Benchmark Extensions}
\label{app:extend}
We have chosen to restrict the Fermi length benchmark to the single-particle and spin-up symmetry subspace. The minimum energy single particle state may not be the minimum energy over all particle numbers. Note that our SWAP-like ansatz and fixed initial state ensure that the ansatz cannot generate any states other than single particle states. Therefore, generation of multi-particle states only occurs due quantum errors, and, in their absence, optimizing over the allowed parameters optimizes over the restricted subspace only. This restricted problem requires fewer quantum resources and has a simple closed form solution that we use to simplify the benchmark evaluation. For the one particle subspace we only require that the top half of the qubits are subjected to gates. This simplification is allowed because the one-particle subspace is made up of two degenerate parts corresponding to spin up and spin down, and either one only requires half of the qubits in our chosen representation. 

In the fully general case, we can segregate any Hamiltonian which has a well defined particle number symmetry into spaces based on these particle numbers. The size of the segregated subspaces naturally obeys the form of $\sum_{m=0}^n \binom{n}{m} = 2^{n}$. Therefore, the full Hilbert space of size $2^n$ is broken up into sets of size $\binom{n}{m}$ for $m\leq n$ particles. It is easy to see that the single particle subspace is linear in the system size $\binom{n}{1}=\mathcal{O}(n)$, while other subspace grow much faster. Notably, the largest subspace is given by the so-called half filling case $m=n/2$, which grows exponentially in $n$.

The subspace size gives us a rough characterization of the classical computational complexity if the solution method is assumed to be brute-force matrix diagonalization within each subspace. Using sparse methods to diagonalize a size $2^n \times 2^n$ matrix within a smaller block of size $\mathcal{O}(n^m)$ takes classical resources (time/memory) that scale polynomially in $n$ for a fixed particle-number subspace $m$ and exponentially as $m$ increases, for example if we fix $m=n/2$.

For each choice of subspace, we can also quantify the minimum number of parameters required to fully specify a quantum state within this space. The single particle subspace can be written as a linear combination of single particle states defined by 
\[
\ket{s}=\alpha\ket{10\cdots0}+\beta\ket{01\cdots0}+\dots + \zeta\ket{00\cdots1}.
\]
Therefore any real state of the above form can be fully specified with a minimal parameter count of $\binom{n}{m}-1=\binom{n}{1}-1=n-1$ real parameters (minus one due to the normalization requirement) . We are not aware of a quantum ansatz which is capable of matching this minimal parameter count in general.  However, an ansatz for less trivial spaces remains an active area of research. Further symmetries of a given Hamiltonian may allow further reductions of these parameter counts, in principle.

If we extend the benchmark to the two particle case and a 2-D lattice structure, there are now two non-degenerate spin subspaces: one where both particles have the same spin (e.g., $s_z=\pm1$) and one where they have opposite spins ($s_z=0$). In the $s_z=1$ case, we still only require gates on the top half of the qubits, but the $s_z=0$ case requires gates on all qubits.

Results from previous investigations on small problems in quantum chemistry show that a similar ansatz performs well in this two particle subspace~\cite{gard2019efficient,doi.org/10.1002/qute.202100012}. Therefore we can extend the protocol used for the benchmark to evaluation of symmetry subspaces for which an exact classical solution is not necessarily known. Because we want to make this extension straightforward, we insist on using the entire space, and not just the spin-up subspace, in our benchmark protocol. The cost of this is that the benchmark requires twice as many qubits as are truly necessary to solve the restricted problem. This choice is typical of the tradeoff between simplification and application-specificity that must be negotiated for any benchmark.

\section{SPAM Mitigation}
\label{app:symmerr}
In order to mitigate errors that occur during state preparation and measurement, we employ a simple construction that maps ideal, intended states to noisy output states using the equation,
\begin{equation}
    M_{ij}x_k=y_k ,
\end{equation}
where $M_{ij}$ is a scattering matrix relating the ideal input states, $x_k$, to their measured noisy output states, $y_k$~\cite{qiskit_spam}. Constructing this scattering matrix without restrictions would require measuring independently all $2^N$ input bit strings for a $N$ qubit system. This is possible, and very effective, for small systems, but becomes untenable rapidly as the number of qubits exceeds $~10$. However, we can reduce this overhead by assuming that SPAM errors for each qubit are uncorrelated with other qubits. Under this assumption, we only need to prepare two measurement circuits, the all-zero state and the all-one state, independent of the number of qubits in the system. The results allow us to populate $N$ sets of size $2 \times 2$ scattering matrices $M_k$. We then construct the full scattering matrix by $M_{ij}=M_1\bigotimes M_2\bigotimes \cdots M_N$. 

For SPAM mitigation, we intend to relate the noisy output states back to ideal input states, therefore we need to calculate $M_{ij}^{-1}$ (or its psuedo-inverse in the case of singular values). For the factorized case described above this inversion is not a challenging task. However, actually carrying out the $M_{ij}^{-1} y_k$ operation is a large calculation in our implementation, even though we avoid explicitly constructing the full $2^N \times 2^N$ matrix. This calculation is the current bottleneck in our classical post-processing and limits SPAM mitigation to problems with less than 26 qubits.

\begin{figure*}
    \centering
    \includegraphics[scale=0.4]{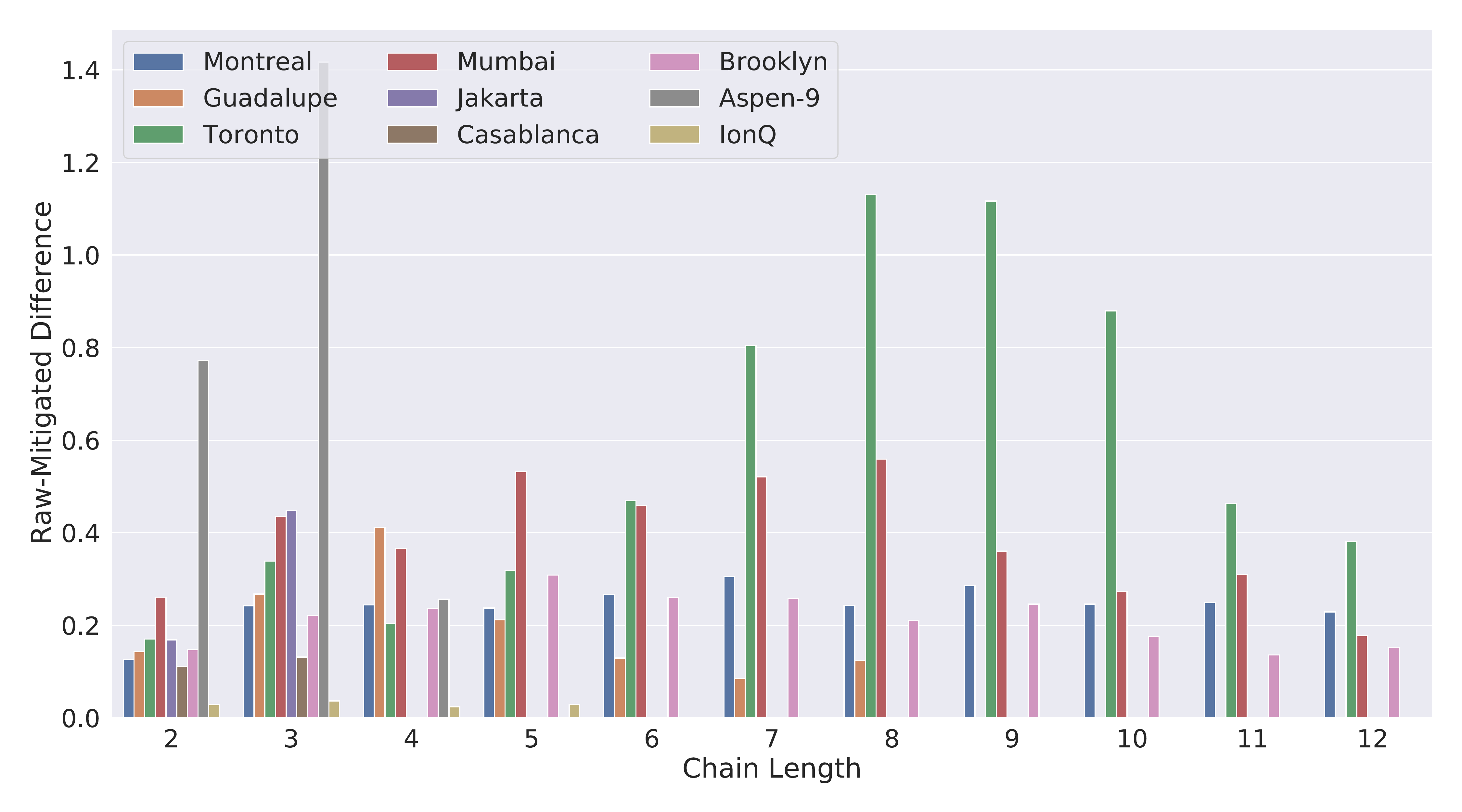}
    \caption{Energy differences between each device's raw results and its SPAM mitigated results. This difference characterizes the magnitude of errors which arise due to uncorrelated SPAM errors. Large, positive differences indicate that the mitigation results in a largely improved energy measurement after mitigation. Not all devices are present at each chain length due to varying physical system size.}
    \label{fig:rawmitigateddiff}
\end{figure*}

To further illustrate the impact of the technique, we show in Fig.~\ref{fig:rawmitigateddiff} the energy correction (difference) between our raw results and the SPAM mitigated  results. The figure captures some information about the nature of the errors on the devices; for example the IonQ device corrections are much smaller than for the Rigetti device, and this is consistent with the difference in self-reported measurement error rates between the two devices.  We also see that there is a significant component of the error in each case that our mitigation does not remove.  These include gate errors and SPAM errors due to correlating or drifting conditions.
 
\end{document}